\PassOptionsToPackage{table,xcdraw}{xcolor}

\documentclass{vgtc}                          

\graphicspath{{figures/}{pictures/}{images/}{./}} 

\usepackage{times}                     

\usepackage{tabu}                      
\usepackage{booktabs}                  
\usepackage{lipsum}                    
\usepackage{mwe}      

\usepackage{mathptmx}                  

\onlineid{1233}

\vgtccategory{Research}

\vgtcinsertpkg

\title{Hands vs. Controllers: Comparing User Interactions in Virtual Reality Shopping Environments}

\author{
Francesco Vona\thanks{e-mail: francesco.vona@hshl.de}\\ %
    \scriptsize Hochschule Hamm-Lippstadt %
    \and 
Julia Schorlemmer\thanks{e-mail: julia.schorlemmer@hshl.de}\\ %
     \scriptsize Hochschule Hamm-Lippstadt %
     \and 
Jessica Stemann\thanks{e-mail: jessica.stemann@hshl.de}\\ %
    \scriptsize Hochschule Hamm-Lippstadt %
    \and
Sebastian Fischer \thanks{e-mail: sebastian.fischer@hshl.de}\\ %
    \scriptsize Hochschule Hamm-Lippstadt %
    \and 
Jan-Niklas Voigt-Antons \thanks{jan-niklas.voigt-antons@hshl.de} \\ %
    \scriptsize Hochschule Hamm-Lippstadt %
}

\abstract{
Virtual reality (VR) enables users to experience real-life situations in immersive environments. Interaction methods significantly shape user experience, particularly in high-fidelity simulations mimicking real-world tasks. This study evaluates two primary VR interaction techniques—hand-based and controller-based—through virtual shopping tasks in a simulated supermarket with 40 participants. Hand-based interaction was preferred for its natural, immersive qualities and alignment with real-world gestures but faced usability challenges, including limited haptic feedback and grasping inefficiencies. In contrast, controller-based interaction offered greater precision and reliability, making it more suitable for tasks requiring fine motor skills.
} 

\renewcommand\copyrighttext{%
  \footnotesize \textcopyright 2025 IEEE. Personal use of this material is permitted. Permission from IEEE must be obtained for all other uses, in any current or future media, including reprinting/republishing this material for advertising or promotional purposes, creating new collective works, for resale or redistribution to servers or lists, or reuse of any copyrighted component of this work in other works. DOI and link to original publication will be added as soon as they are available.}
  
\newcommand\copyrightnotice{%
   \fbox{\parbox{\dimexpr\columnwidth-\fboxsep-\fboxrule\relax}{\copyrighttext}}
}


\begin{document}

\maketitle
\copyrightnotice
\section{Introduction and Related work}
Virtual Reality (VR) technology enables immersive experiences in gaming, training, healthcare, and education by simulating real-world environments. Besides visual quality, interaction methods, such as hand-based and controller-based approaches, are crucial, as they significantly impact user experience and task performance in VR settings. Hand-based methods mimic natural gestures for intuitive and immersive interactions, while controllers offer precision and tactile feedback, making them suitable for tasks requiring accuracy. While research highlights the importance of interaction techniques, there is limited empirical data comparing hand-based and controller-based interactions in high-fidelity simulations like training or industrial tasks \cite{b1, b2, b3}. These scenarios often demand both natural interaction and precision, emphasizing trade-offs between immersion and control. This study examines these methods in a virtual supermarket to evaluate usability and user experience. Studies already highlighted the naturalness and immersion of hand-tracking, which enhances users’ sense of embodiment by eliminating the need for physical controllers. 
However, challenges such as reduced precision, latency, and the absence of haptic feedback limit its performance, particularly in tasks requiring fine motor skills \cite{b5, b6}. Controllers consistently outperform hand-tracking in accuracy and task speed, making them reliable for precision-based activities \cite{b6}. In gaming, research has shown that controllers excel in action-intensive tasks while hand-tracking is favored in narrative-driven or exploratory contexts due to its immersive qualities \cite{b8, b7}. However, some studies challenge the assumption that hand-tracking always improves realism, finding no significant increase in perceived naturalness during object manipulation \cite{b9}. On the other hand, the importance of haptic feedback is well-documented, with studies demonstrating that even basic tactile cues enhance user experience and task performance \cite{b12, b13}. In VR contexts requiring object manipulation—such as virtual shopping—controllers are preferred for their precision, though hand-tracking’s intuitive appeal makes it popular for more immersive applications \cite{b14, b8}. Despite numerous conflicting studies regarding preferences between hands or controllers in various contexts, this research seeks to ascertain the preferred method within an immersive shopping experience, where precision and sense of immersion are key considerations.
\vspace{-0.4cm}
\section{Methods}
The objective of this experiment was to evaluate the usability, user preferences, and user experience of two interaction methods—hand-tracking and controllers—within a virtual supermarket environment. The study was designed as a within-subjects experiment, meaning that all participants experienced both conditions: hand-based interaction (Condition A) and controller-based interaction (Condition B). This design allowed for a direct comparison between the two methods for each participant. The study received approval from the university's ethics committee. Participants performed two tasks, each requiring them to select three cereal packages from a designated category. In Task 1, participants selected “healthy” cereal packages, while in Task 2, they chose “tasty” cereals. Both tasks involved interacting with a virtual supermarket shelf. The order of tasks and interaction methods was randomized for each participant, ensuring a balanced approach. Each participant completed both conditions once. To minimize visual differences between the hand-tracking and controller-based methods, participants’ virtual hands were represented identically for both interaction techniques. In Condition A (hand-based interaction), participants mimicked a natural grasping motion to pick up the cereal packages. In contrast, in Condition B (controller-based interaction), they used a VR controller and pressed the trigger button to interact with the items. In both conditions, participants placed the selected items into a virtual shopping cart by releasing their grasp or the controller’s trigger button, depending on the interaction method. No haptics feedback was provided for Condition B.
Collected data included: i) Demographic Data, including age, gender, and occupation; ii) Affinity for Technology Interaction (ATI) \cite{b15}; iii) System Usability Scale (SUS) \cite{b16}; iv) User Experience Questionnaire Short (UEQ-S) \cite{b17} and v) Custom Questions concerning the interaction method preferences and the experienced ease of use through a 7 points Likert-scale. The study took place in a controlled laboratory environment, with each session lasting about 30 minutes. The sample included 40 participants (18 male, 22 female) aged 19-32 years (M = 22.1, SD = 3.24), predominantly university students (90\%). The ATI scale indicated a moderate affinity for technology (M = 3.41, SD = 0.96), and the scale's internal consistency was strong ($\alpha = 0.89$).
\begin{figure}[ht]
    \centering
    \includegraphics[alt={Bar-chart showing the preferred and easiest interaction method among participants divided by gender.}, width=0.95\columnwidth]{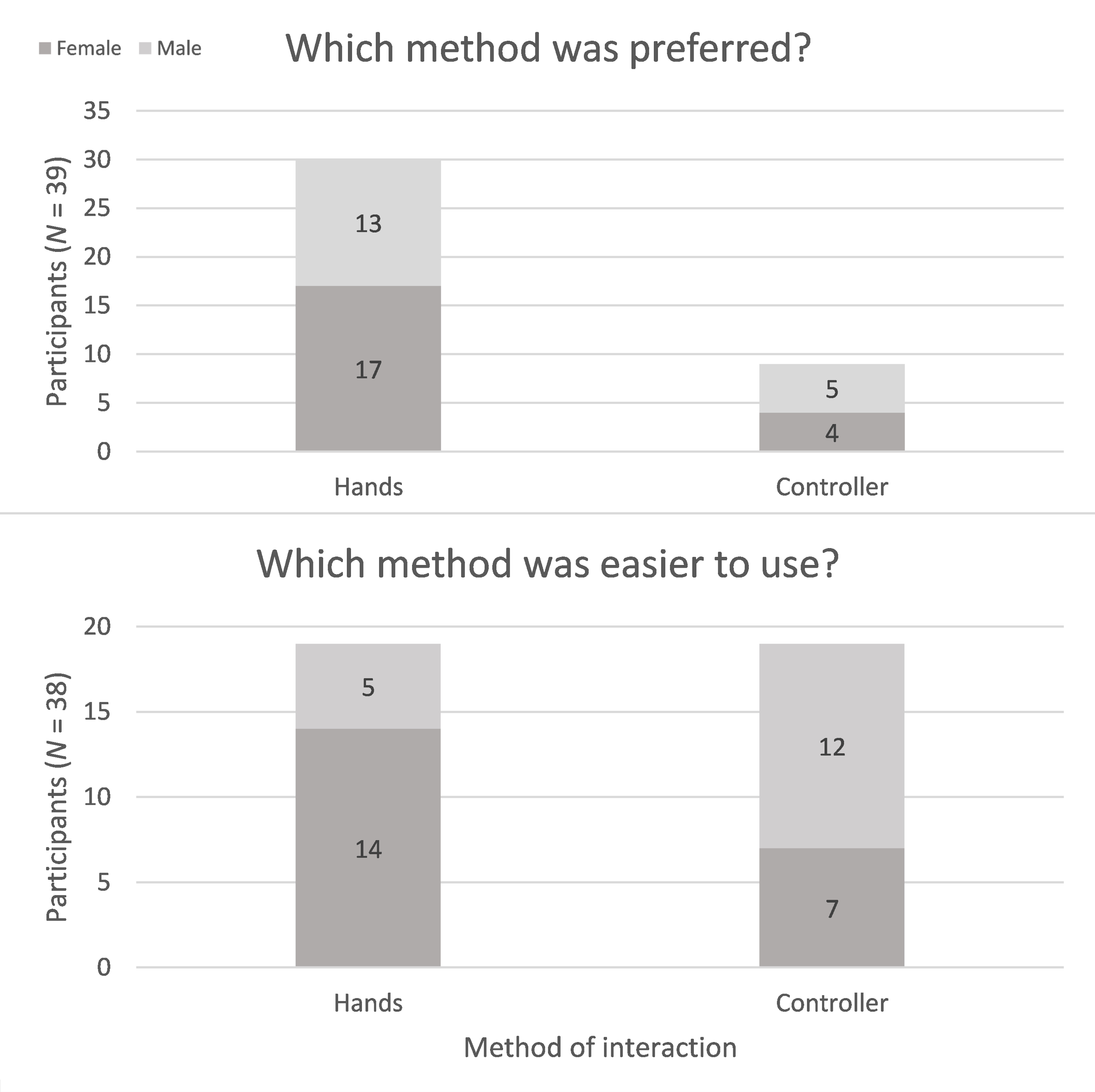}
    \caption{Preferred and easiest interaction method among participants divided by gender.}
    \label{fig:methodpreferred}
\end{figure}
\vspace{-0.35cm}
\section{Results and Discussion}
The User Experience Questionnaire (UEQ) revealed similar Pragmatic Quality scores for both interaction methods, indicating participants found both methods functional for task completion (hand-based M = 1.61, controllers M = 1.65). Hedonic Quality, measuring engagement and enjoyment, slightly favored hand-based interaction (M = 2.00) over controllers (M = 1.94). The System Usability Scale (SUS) scores were comparable, with hand-based scoring M = 71.38 and controllers M = 68.88. Both scores (UEQ and SUS) suggest both methods were perceived as similarly usable, although the differences were not statistically significant. 75\% of participants preferred the immersive nature of hand-based interaction, aligning with its natural and intuitive qualities (Figure \ref{fig:methodpreferred}). However, usability challenges, such as difficulty accurately grasping objects due to the lack of haptic feedback, hindered its effectiveness. Ease of use was evenly split, with 48\% of participants favoring each method (Figure \ref{fig:methodpreferred}). Female participants found hand-based interaction easier to use, while male participants preferred controllers, possibly due to greater familiarity with gaming controllers. This suggests prior exposure to input devices may influence preferences \cite{b1}. Hand-based interaction often resulted in multiple attempts to grasp and manipulate objects, causing frustration and inefficiencies. In contrast, controllers provided greater precision, enabling smoother object manipulation. Hand-based interaction enhanced immersion and user engagement due to its alignment with real-world gestures, consistent with prior research on natural interfaces \cite{b14}. However, these benefits did not translate into better task performance \cite{b1}. Challenges such as the absence of haptic feedback and reduced accuracy limited usability for precision tasks \cite{b12}. A lack of tactile feedback was a significant limitation for hand-based interaction, reducing task efficiency and increasing error rates. Prior studies have shown that even minimal haptic cues can improve object manipulation and overall user experience \cite{b3}. Incorporating such feedback into hand-based systems could address these shortcomings. Both methods were rated highly in functionality and engagement, with hand-based interaction slightly outperforming controllers in Hedonic Quality. This aligns with the preference for natural interactions in VR environments designed for exploratory or slower-paced tasks \cite{b3}. Unfortunately, not all participants answered the custom questions (39 for the preferred method, 38 for the ease of use).
\vspace{-0.3cm}
\section{Conclusion}
The findings emphasize the need for customizable interaction methods to cater to diverse user preferences. For instance, integrating optional haptic feedback and allowing users to toggle between interaction modes based on task requirements could improve VR usability. While hand-based interaction was preferred for its immersive qualities, limitations such as the lack of haptic feedback and grasping difficulties impacted usability. Controller-based interaction, though less immersive, offered greater precision and efficiency in the supermarket environment. Future studies should explore more complex tasks to deepen insights into usability and performance in VR systems, possibly with a larger sample of users.



\end{document}